\documentclass{article}
\usepackage{amssymb}
\usepackage{amsmath}
\usepackage{amsfonts}

\input{tcilatex}
\begin{document}

\title{Effect of a possible cosmological time dependence of the
gravitational parameter $G$ on the peak luminosity of type Ia supernovae.}
\author{A. Tartaglia and N. Radicella \\
Dipartimento di Fisica, Politecnico, and INFN, Torino, Italy\\
angelo.tartaglia@polito.it, ninfa.radicella@polito.it}
\maketitle

\begin{abstract}
The cosmological expansion of the universe affects the behaviour of all
physical systems and, in the case of gravitationally bound ones, could
correspond to or mimic a time dependent Newton's constant. Here we discuss
the case of a locally spherical mass distribution embedded in a generic
Robertson Walker universe. Choosing the most appropriate metric tensor for
the problem and assuming that the local time scale is much much lower than
the cosmic one, we show that $G$ is practically unaffected thus leaving the
absolute magnitude of type Ia supernovae unaltered at all epochs.
\end{abstract}

\section{\protect\bigskip Introduction}

An open and since long time debated problem is the one of the effect of the
cosmic expansion on the behavior of gravitating systems on a more or less
local scale (a few examples out of the wide litterature are \cite{McVittie} 
\cite{einstein} \cite{dicky} \cite{petrosian} \cite{cooper} \cite{Citazioni
varie}). Of the various ways this problem can affect the understanding of
our universe, we would like to focus on the collapse of a white dwarf
leading to a supernova of type Ia \cite{snIa} \cite{snIa1} (SnIa). SnIa's
are particularly important because they have led to the discovery of the
accelerated expansion of the universe. The fundamental feature of SnIa's for
the monitoring of the expansion is their being considered as "standard
candles". The latter conviction is based on the fact that the peak
luminosity of this kind of supernova is proportional to the mass of nickel
synthesized during the collapse, which, to a good approximation, is a fixed
fraction of the Chandrasekhar mass \cite{nickel}. The Chandrasekhar mass, in
its turn, is proportional to $G^{-3/2}$ \cite{varG} where $G$ is the
universal coupling constant between matter and geometry (Newton's constant).

There are alternative theories to General Relativity (GR) for which $G$ is
no constant at all, but rather it depends on cosmic time (examples are \cite%
{jordan} \cite{brans} \cite{bergmann} \cite{wagoner} \cite{moffat} \cite%
{maeda} \cite{Gvariabile}). In that case, provided one has the explicit time
dependence of "$G$", the absolute luminosity to be attributed to an SnIa
should appropriately be corrected on the basis of the corresponding redshift 
$z$ (distance in time) in order to separate the contribution of the
expansion from the one of the changing $G$.

Even in the case of GR (possibly "extended" \cite{CD}), however, one could
surmise an influence of the expansion of the universe appearing in the form
of an effective $G_{eff}$, i.e. something "dressing" the universal $G$ with
a time- or $z$-dependence of the local gravitational interaction.

In any case the presence of a time ($z$) dependent $G_{eff}$ (whatever be
the origin of the dependence) would imply a correction on the absolute
magnitude $M\left( z\right) $ of an SnIa, given by the formula \cite%
{trequarti}:%
\begin{equation*}
M\left( 0\right) -M\left( z\right) =\frac{15}{4}\log \frac{G_{eff}\left(
z\right) }{G_{eff}\left( 0\right) }
\end{equation*}

Considering the relevance of this problem, in the present letter we shall
focus on the supernova luminosity in any Robertson Walker (RW) spacely flat
universe, i.e. any expanding universe in which the gravitational interaction
is expressed by the geometry of a four dimensional manifold with a global
symmetry reducing the number of independent functions, in the average cosmic
metric tensor, to the only scale factor $a$.

We shall start from an ansatz for the metric describing the situation around
a local spherical distribution of mass in an expanding RW universe. Then
exploiting the fact that a supernova explosion is an extremely short event
on the scale of cosmic times we shall verify that the implied $G$ value is
in any case the universal one.

\section{A spherically symmetric perturbation in a Robertson Walker universe}

Let us start from the remark that at the scale of a stellar system the
effect of the cosmic curvature appears only as a very tiny perturbation of
the usual stationary state solutions of general relativity. Considering for
instance the example of a spherically distributed bunch of matter in a
background RW universe with zero space curvature, we may guess and assume
that locally the induced metric is essentially 
\begin{equation}
ds^{2}=\left( 1-f\left( r,\tau \right) \right) d\tau ^{2}-a^{2}\left( \tau
\right) \left( \frac{dr^{2}}{1-h\left( r,\tau \right) }+r^{2}d\theta
^{2}+r^{2}\sin ^{2}\theta d\phi ^{2}\right)  \label{lineanewton}
\end{equation}%
where $a$ is the cosmological scale factor (adimensional) and an extremely
weak dependence of $f$ and $h$ on time is expected. Our $\tau $ is measured
in meters.

Far away from the local source it must be $f$, $h\rightarrow 0$ so that the
pure RW metric is recovered. On the other side if it were $a=a_{0}=$
constant we could incorporate its value in the definition of $r$ and from
the field equations we would obtain the Schwarzschild solution $%
f=h=2Gm/c^{2}r=2\mu /r$.

A metric like the one in (\ref{lineanewton}) belongs to a general class of
exact solutions of the Einstein equations discussed for instance in \cite%
{stephani}. When applyed to a cosmological problem it has a partial
correspondence with the early work of McVittie \cite{McVittie}, many times
reconsidered and discussed afterwards (see for instance \cite{franca} \cite%
{nolan}). Actually, however, he analyzed a spherical massive source immersed
in a homogeneous fluid with which it interacted gravitationally. Our case
(and specific treatment) is different because for us the cosmic fluid is
implicitly unaffected by the local spherical source, which assumtion is
indeed realistic being the cosmic fluid itself the product of an average
over the whole universe. What we want to elucidate is the inverse, i.e. the
effect of the global (average) curvature on the local system.

Without loosing generality we may restyle (\ref{lineanewton}) evidencing an
arbitrary moment in cosmic time. The corresponding $a\left( \tau _{0}\right)
=a_{0}$ (the value of the scale parameter at the chosen moment) may be
absorbed into a rescaling of $r$, and we introduce the new renormalized
scale factor $\alpha \left( \tau ,\tau _{0}\right) =a\left( \tau \right)
/a\left( \tau _{0}\right) $.

From the line element written in this form we can compute the Einstein
tensor $G_{\mu \nu }$, assuming that $f$, $h<<1$ which is an obvious
situation when black holes are not implied (actually, in "ordinary"
conditions, $f$, $h$ $<\sim 10^{-6}$):%
\begin{eqnarray}
G_{00} &=&3\frac{\dot{\alpha}^{2}}{\alpha ^{2}}-\frac{1}{r\alpha ^{2}}\left(
h^{\prime }+\frac{h}{r}\right) -\frac{\dot{\alpha}}{\alpha }\dot{h}  \notag
\\
G_{rr} &=&\frac{1}{r}\left( f^{\prime }+\frac{h}{r}\right) +\alpha \dot{%
\alpha}\dot{f}+\left( 2\alpha \ddot{\alpha}+\dot{\alpha}^{2}\right) \left(
f+h-1\right)  \notag \\
G_{\theta \theta } &=&\left( 2\alpha \ddot{\alpha}+\dot{\alpha}^{2}\right)
r^{2}\left( f-1\right) +\frac{1}{2}\alpha ^{2}r^{2}\ddot{h}  \label{einstein}
\\
&&+\alpha \dot{\alpha}r^{2}\left( \dot{f}+\frac{3}{2}\dot{h}\right) +\frac{1%
}{2}r^{2}f^{\prime \prime }+\frac{1}{2}r\left( f^{\prime }+h^{\prime }\right)
\notag \\
G_{\phi \phi } &=&G_{\theta \theta }\sin ^{2}\theta  \notag \\
G_{0r} &=&\frac{\dot{\alpha}}{\alpha }f^{\prime }-\frac{\dot{h}}{r}  \notag
\end{eqnarray}%
Primes mean partial derivatives with respect to $r$ and dots correspond to
partial $\tau $ derivatives.

One further assumption is that the source term in the Einstein equations can
be decomposed into a cosmic contribution $T_{\mu \nu }$ (peculiar to the
given model one is considering) and a local contribution $\mathfrak{T}_{\mu
\nu }$, so that the equations can be written:%
\begin{equation*}
G_{\mu \nu }=\kappa T_{\mu \nu }+\kappa \mathfrak{T}_{\mu \nu }.
\end{equation*}%
The two contributions are necessarily very different from one another and
the second is locally much bigger than the first. Having assumed, on the
average, a RW symmetry, we can verify, by direct inspection of (\ref%
{einstein}), that the terms appearing in the expression of $G_{\mu \nu }$
not containing $f$ and $h$ correspond to the cosmic source, so that the
remaining ones pertain to the local source. If we consider the situation
outside the local distribution of matter we are in "vacuo", so that the
additional terms of (\ref{einstein}) give rise to the equations%
\begin{gather}
\frac{1}{r\alpha ^{2}}\left( h^{\prime }+\frac{h}{r}\right) +\frac{\dot{%
\alpha}}{\alpha }\dot{h}=0  \notag \\
\frac{1}{r}\left( f^{\prime }+\frac{h}{r}\right) +\alpha \dot{\alpha}\dot{f}%
+\left( 2\alpha \ddot{a}+\alpha ^{2}\right) \left( f+h\right) =0  \notag \\
\left( 2\alpha \ddot{a}+\dot{\alpha}^{2}\right) r^{2}f+\frac{1}{2}\alpha
^{2}r^{2}\ddot{h}+\alpha \dot{\alpha}r^{2}\left( \dot{f}+\frac{3}{2}\dot{h}%
\right)  \label{equazioni} \\
+\frac{1}{2}r^{2}f^{\prime \prime }+\frac{1}{2}r\left( f^{\prime }+h^{\prime
}\right) =0  \notag \\
\frac{\dot{\alpha}}{\alpha }f^{\prime }-\frac{\dot{h}}{r}=0  \notag
\end{gather}

The assumed weak dependence on time allows for a low order power series
development around the local (in time) expression of the functions.
Introducing $\mathfrak{t}=\tau -\tau _{0}$ as the time variable we write: 
\begin{eqnarray}
\alpha \left( \mathfrak{t}\right) &=&1+H_{0}\mathfrak{t-}\frac{1}{2}%
q_{0}H_{0}^{2}\mathfrak{t}^{2}+...  \notag \\
f\left( r,\mathfrak{t}\right) &=&f_{0}\left( r\right) +f_{1}\left( r\right) 
\mathfrak{t+}f_{2}\left( r\right) \mathfrak{t}^{2}+...  \label{serie} \\
h\left( r,\mathfrak{t}\right) &=&h_{0}\left( r\right) +h_{1}\left( r\right) 
\mathfrak{t+}h_{2}\left( r\right) \mathfrak{t}^{2}+...  \notag
\end{eqnarray}%
Just to fix the orders of magnitude let us remark that:%
\begin{eqnarray*}
f,h &\lesssim &10^{-6}\text{ at most} \\
H_{0} &\sim &10^{-18}\text{ s}^{-1}\doteqdot \sim 10^{-26}\text{ m}^{-1}
\end{eqnarray*}%
The Hubble parameter correction can produce a contribution comparable to the
one from $f$ and $h$ in times of the order of $\sim 10^{5}$ years.

Introducing (\ref{serie}) into (\ref{equazioni}) and fixing the attention on
the zeroth order in $\mathfrak{t}$ we obtain:%
\begin{gather}
\frac{h_{0}^{\prime }}{r}+\frac{h_{0}}{r^{2}}+H_{0}h_{1}=0  \notag \\
\frac{f_{0}^{\prime }}{r}+\frac{h_{0}}{r^{2}}+H_{0}^{2}\left(
1-2q_{0}\right) \left( f_{0}+h_{0}\right) +H_{0}f_{1}=0  \notag \\
h_{2}r^{2}+H_{0}r^{2}\left( f_{1}+\frac{3}{2}h_{1}\right) +\frac{r^{2}}{2}%
f_{0}^{\prime \prime }+H_{0}^{2}\left( 1-2q_{0}\right) r^{2}f_{0}
\label{zero} \\
+\frac{r}{2}\left( h_{0}^{\prime }+f_{0}^{\prime }\right) =0  \notag \\
H_{0}f_{0}^{\prime }-\frac{h_{1}}{r}=0  \notag
\end{gather}%
The system (\ref{zero}) is easily solved, starting from the ansatz: 
\begin{equation*}
f_{0}=2\frac{\mu }{r}.
\end{equation*}

\bigskip The solution is%
\begin{eqnarray}
f_{0} &=&2\frac{\mu }{r}  \notag \\
h_{0} &=&\frac{A}{r}+H_{0}^{2}\mu r  \notag \\
f_{1} &=&-\frac{A-2\mu }{H_{0}r^{3}}-H_{0}\frac{\mu }{r}-H_{0}\left(
1-2q_{0}\right) \left( \frac{2\mu +A}{r}+H_{0}^{2}\mu r\right)
\label{soluzioni} \\
h_{1} &=&-2H_{0}\frac{\mu }{r}  \notag \\
h_{2} &=&\frac{3}{r^{3}}\left( \frac{A}{2}-\mu \right) +\frac{A}{r}%
H_{0}^{2}\left( 1-2q_{0}\right) +\frac{H_{0}^{2}}{r}\frac{7}{2}\mu +r\mu
H_{0}^{4}\left( 1-2q_{0}\right)  \notag
\end{eqnarray}%
where $A$ is an integration constant.

In terms of the full functions we would write

\begin{eqnarray*}
f\left( r,\mathfrak{t}\right) &=&2\frac{\mu }{r}-\left( \frac{A-2\mu }{%
H_{0}r^{3}}+H_{0}\frac{\mu }{r}+H_{0}\left( 1-2q_{0}\right) \left( \frac{%
2\mu +A}{r}+H_{0}^{2}\mu r\right) \right) \mathfrak{t+...} \\
h\left( r,\mathfrak{t}\right) &=&\frac{A}{r}+H_{0}^{2}\mu r-2H_{0}\frac{\mu 
}{r}\mathfrak{t} \\
&&\mathfrak{+}\left( \frac{3}{r^{3}}\left( \frac{A}{2}-\mu \right) +\frac{A}{%
r}H_{0}^{2}\left( 1-2q_{0}\right) +\frac{H_{0}^{2}}{r}\frac{7}{2}\mu +r\mu
H_{0}^{4}\left( 1-2q_{0}\right) \right) \mathfrak{t}^{2}+...
\end{eqnarray*}%
In a flat, static background it would be $H_{0}=0$ and the solution should
coincide with Schwarzschild's, so that it must be%
\begin{equation*}
A=2\mu
\end{equation*}%
and%
\begin{equation*}
f\left( r,\mathfrak{t}\right) =2\frac{\mu }{r}-\mu \left( \frac{H_{0}}{r}%
+H_{0}\left( 1-2q_{0}\right) \left( \frac{4}{r}+H_{0}^{2}r\right) \right) 
\mathfrak{t+...}
\end{equation*}%
\begin{multline}
h\left( r,\mathfrak{t}\right) =\frac{2\mu }{r}+H_{0}^{2}\mu r-2H_{0}\frac{%
\mu }{r}\mathfrak{t}  \label{serie11} \\
\mathfrak{+\mu }H_{0}^{2}\left( rH_{0}^{2}\left( 1-2q_{0}\right) +\frac{11}{%
2r}-4\frac{q_{0}}{r}\right) \mathfrak{t}^{2}+...  \notag
\end{multline}

We see that the expansion of the universe shows up in a "decoupling" of $h$
from $f$ already at the zero order. However this correction is extremely
small. In fact the zero order for $h$ is:%
\begin{equation*}
\frac{2\mu }{r}\left( 1+\frac{H_{0}^{2}}{2}r^{2}\right)
\end{equation*}%
and the difference from the typical Schwarzschild term appears ($1\%$
correction) at distances $\sim 10^{25}$ m $\doteqdot 10^{3}$ Mpc!

\section{Behavior of test particles}

In order to investigate further on the possible local effects of the
expanding universe it is convenient to write down the geodesics for the
metric in (\ref{lineanewton}). Directly exploiting the spherical symmetry,
which is not spoiled by the expansion, we may fix $\theta =\pi /2$, so that
the three remaining independent equations for the motion of a test particle
are

\begin{multline*}
\frac{d^{2}\tau }{ds^{2}}+\alpha \dot{\alpha}\left( \frac{dr}{ds}\right)
^{2}+\alpha \dot{\alpha}r^{2}\left( \frac{d\phi }{ds}\right) ^{2}-\frac{\dot{%
f}}{2}\left( \frac{d\tau }{ds}\right) ^{2}+f^{\prime }\frac{dr}{ds}\frac{%
d\tau }{ds} \\
-\left( \frac{\dot{h}}{2}\alpha ^{2}+\alpha \dot{\alpha}\left( h+f\right)
\right) \left( \frac{dr}{ds}\right) ^{2}-\alpha \dot{\alpha}fr^{2}\left( 
\frac{d\phi }{ds}\right) ^{2}=0
\end{multline*}%
\begin{multline}
\frac{d^{2}r}{ds^{2}}+2\frac{\dot{\alpha}}{\alpha }\frac{dr}{ds}\frac{d\tau 
}{ds}-r\left( \frac{d\phi }{ds}\right) ^{2}-\frac{f^{\prime }}{2\alpha ^{2}}%
\left( \frac{d\tau }{ds}\right) ^{2}  \label{erre} \\
-\dot{h}\frac{dr}{ds}\frac{d\tau }{ds}-\frac{h^{\prime }}{2}\left( \frac{dr}{%
ds}\right) ^{2}-hr\left( \frac{d\phi }{ds}\right) ^{2}=0  \notag
\end{multline}%
\begin{equation}
\frac{d^{2}\phi }{ds^{2}}+\frac{2}{r}\frac{dr}{ds}\frac{d\phi }{ds}+2\frac{%
\dot{\alpha}}{\alpha }\frac{d\phi }{ds}\frac{d\tau }{ds}=0  \label{phi}
\end{equation}%
Eq. (\ref{phi}) is easily integrated to yield:%
\begin{equation}
\frac{d\phi }{ds}=\frac{L}{\alpha ^{2}r^{2}}  \label{elle}
\end{equation}%
being $L$ a constant of the motion (angular momentum).

Using (\ref{elle}) the remaining two equations read:%
\begin{equation}
\left\{ 
\begin{array}{c}
\begin{array}{c}
\frac{d^{2}\tau }{ds^{2}}+\alpha \dot{\alpha}\left( \frac{dr}{ds}\right)
^{2}+\alpha \dot{\alpha}r^{2}\left( 1-f\right) \left( \frac{L}{\alpha
^{2}r^{2}}\right) ^{2} \\ 
-\frac{\dot{f}}{2}\left( \frac{d\tau }{ds}\right) ^{2}+f^{\prime }\frac{dr}{%
ds}\frac{d\tau }{ds}-\left( \frac{\dot{h}}{2}\alpha ^{2}+\alpha \dot{\alpha}%
\left( h+f\right) \right) \left( \frac{dr}{ds}\right) ^{2}=0%
\end{array}
\\ 
\begin{array}{c}
\frac{d^{2}r}{ds^{2}}+2\frac{\dot{\alpha}}{\alpha }\frac{dr}{ds}\frac{d\tau 
}{ds}-r\left( 1+h\right) \left( \frac{L}{\alpha ^{2}r^{2}}\right) ^{2} \\ 
-\frac{f^{\prime }}{2\alpha ^{2}}\left( \frac{d\tau }{ds}\right) ^{2}-\dot{h}%
\frac{dr}{ds}\frac{d\tau }{ds}-\frac{h^{\prime }}{2}\left( \frac{dr}{ds}%
\right) ^{2}=0%
\end{array}%
\end{array}%
\right.  \label{due}
\end{equation}

We may also directly obtain $dr/ds$ from the line element (\ref{lineanewton}%
):%
\begin{equation}
\left( \frac{dr}{ds}\right) ^{2}=\frac{\left( 1-f\right) \left( 1-h\right) }{%
\alpha ^{2}}\left( \frac{d\tau }{ds}\right) ^{2}-\frac{1-h}{\alpha ^{2}}-%
\frac{L^{2}}{\alpha ^{4}r^{2}}\left( 1-h\right)  \label{erre1}
\end{equation}

The last three equations are what is needed to pursue any further analysis
of the behaviour of matter in our system.

\subsection{A special case: a fixed object}

We consider now a little object maintained at a fixed position $r=R$ in the
field (by means of some appropriate force); the angular momentum will then
be $L=0$. In this case equations (\ref{due}) become:%
\begin{equation}
\left\{ 
\begin{array}{c}
\frac{d^{2}\tau }{ds^{2}}-\frac{\dot{f}}{2}\left( \frac{d\tau }{ds}\right)
^{2}=\mathfrak{g}_{0} \\ 
\frac{\left. f^{\prime }\right\vert _{r=R}}{2\alpha ^{2}}\left( \frac{d\tau 
}{ds}\right) ^{2}=-\mathfrak{g}_{r}%
\end{array}%
\right.  \label{statico}
\end{equation}%
where $\mathfrak{g}$ is the four-vector representing the force per unit mass
acting upon our object.

At this point let us introduce the approximate solutions for $\alpha ,f,h$
and keep the $0$ order in $\mathfrak{t}$:%
\begin{gather*}
\frac{d^{2}\tau }{ds^{2}}+H_{0}\mu \left( 2qR+\frac{R}{2}H_{0}^{2}+8\frac{q}{%
H_{0}^{2}R}+\frac{5}{2R}\right) \left( \frac{d\tau }{ds}\right) ^{2}=%
\mathfrak{g}_{0} \\
\frac{\mu }{R^{2}}\left( \frac{d\tau }{ds}\right) ^{2}=-\mathfrak{g}_{r}
\end{gather*}%
The first equation tells us that, for $r=R$, $\frac{d\tau }{ds}=K\left(
R\right) $ also is a constant (to say it better: it is a function of $R$).

This result is then introduced into the second equation, thus yielding (now $%
g\left( R\right) =c^{2}\frac{\mathfrak{g}\left( R\right) }{K^{2}\left(
R\right) }$ and the explicit expression for $\mu $ has been restored): 
\begin{equation}
-\frac{GM}{R^{2}}=g\left( R\right)  \label{newton}
\end{equation}%
where $g$ is in practice the local gravitational acceleration and of course (%
\ref{newton}) is Newton's law.

What is indeed remarkable is that the coupling constant is the universal $G$%
, unaffected by the expansion of the universe. A phenomenon which happens in
a short time, such as the implosion of a star due to a gravitational
collapse, will be determined and measured by the same value of $G$ at all
epochs. This is of paramount importance, as we noticed before, for the
evaluation of the apparent luminosity of type Ia supernovae, which depends
on $G^{-3/4}$.

\section{Conclusion}

We have studied the case of a spherical distribution of matter embedded in a
RW spacely flat universe. Assuming that the global symmetry and behaviour of
the universe are not affected by the local source, we have analyzed the
influence of the cosmic expansion on the behaviour of the local system. What
we have seen is that, for sufficiently small time spans (zeroth order
expansion in time of the functions in the metric) the three-dimensional
force acting upon a static test mass is indeed expressed by Newton's law
without any epoch dependent renormalization of $G$. In this way the peak
luminosity of SnIa's, which depends on $G^{-3/2}$ and is due to a very rapid
phenomenon, stays the same at all cosmic times. This result turns out to be
valid for any RW type universe, i.e. for all theories where gravity is
described by the curvature of a four-dimensional Riemannian manifold with
isotropic homogeneous (in the mean) sources and flat three-dimensional space
submanifold. This is true for instance for the Friedmann Robertson Walker,
the $\Lambda $ cold dark matter, and the cosmic defect theory \cite{CD}. In
fact the result we have obtained could be anticipated considering that the
gravitational interaction in an Einstein-type theory is always mediated by
the unique coupling constant between matter and geometry, no matters where
you are in cosmic time. Any cosmological effect shows up only on space and
time scales far bigger than the ones of a local system.

\end{document}